\documentstyle[eqsecnum,prd,aps,epsf]{revtex}
\draft
\begin{document}
\newcommand{\beq}{\begin{equation}}
\newcommand{\eeq}{\end{equation}}
\newcommand{\beqn}{\begin{eqnarray}}
\newcommand{\eeqn}{\end{eqnarray}}
\newcommand{\pa}{\partial}

\twocolumn[\hsize\textwidth\columnwidth\hsize\csname
@twocolumnfalse\endcsname

\begin{center}
{\large\bf{A relativistic formalism for computation of irrotational 
binary stars \\ in quasi equilibrium states}}
~\\
~\\
Masaru Shibata\\
{\em Department of Earth and Space Science,~Graduate School of
  Science,~Osaka University,\\
Toyonaka, Osaka 560-0043, Japan}\\
\end{center}
\begin{abstract}
We present relativistic hydrostatic equations for obtaining 
irrotational binary neutron stars in quasi equilibrium states 
in 3+1 formalism. Equations derived here are different from 
those previously given by Bonazzola, Gourgoulhon, and Marck, 
and have a simpler and more tractable form for computation in 
numerical relativity. 
We also present hydrostatic equations for computation of 
equilibrium irrotational binary stars in first post-Newtonian order. 
\end{abstract}
\pacs{04.25.Nx, 97.80.Gm}
\vskip2pc]
\section{Introduction}

Preparation of reliable theoretical models on late inspiraling 
stage of binary neutron stars is one of the most important issues 
for gravitational wave astronomy. This is 
because they are one of promising sources 
for gravitational wave detector such as LIGO\cite{LIGO}, 
VIRGO\cite{VIRGO}, GEO600\cite{GEO} and TAMA\cite{TAMA}. 
{}From their signals, we will get a wide variety of physical 
information on neutron stars such as their mass, spin, and so on 
if we have a theoretical template of them\cite{KIP}. 
In particular, a signal from very late inspiraling stage 
just prior to merging may contain physically important information 
on neutron stars such as their radius\cite{KIP}, which will be 
utilized for determining equation of state of neutron stars\cite{lindblom}. 

Binary neutron stars evolve due to radiation reaction 
of gravitational waves, so that they never settle down to equilibrium 
states. However, the emission time scale will be 
always longer than the orbital period outside their
innermost stable circular orbit (ISCO), 
so that we may consider that they are in quasi equilibrium states 
in their inspiraling phase even near ISCO. 
Motivated by this idea, there have been several works in which 
sequence of equilibrium states of binary neutron stars are computed 
and the sequence is regarded as an evolutionary track; for example, 
we have obtained corotational equilibrium states in first post-Newtonian 
approximation\cite{shiba}; 
Baumgarte et al. have obtained corotational equilibrium states 
in a relativistic frame work using conformal flat approximation\cite{BCSST}. 
Up to now, however, all relativistic works have been done assuming 
corotational velocity field\cite{uryu}. As pointed out 
previously\cite{Cutler}, corotation is not a adequate assumption for 
velocity field of realistic binary neutron stars, 
because effect of viscosity is negligible 
for evolution of neutron stars in binary and as a result, 
their velocity field are expected to be irrotational 
(or nearly irrotational). 

For computation of realistic quasi equilibrium states of 
coalescing binary neutron stars just prior to merging, 
Bonazzola, Gourgoulhon, and Marck (BGM)\cite{BGM} recently 
presented a relativistic formalism. 
In their formulation, they assume a helicoidal Killing vector $\ell^{\mu}$, 
and then project relativistic 
hydrodynamic equations onto a hypersurface orthogonal to $\ell^{\mu}$. 
After that, they impose their irrotational condition on the 
hypersurface and derive hydrostatic equations for the irrotational fluid. 
We think, however, that there were several inadequate treatments 
in their work. 
First one is their definition of irrotational condition, 
because their irrotational condition is nothing but a necessary 
condition for irrotation even in the case 
when we assume existence of $\ell^{\mu}$\cite{HE}. In general case, 
their condition is not identical with the irrotational condition. 
Second, in numerical relativity, we usually solve equations such as 
Hamiltonian constraint, momentum constraint, and 
equations for gauge conditions, using spatial 
coordinates on the hypersurface, $\Sigma_t$, 
which is perpendicular to unit normal $n^{\nu}$. 
Due to this reason, they had to re-project 
their equations onto $\Sigma_t$. As a result, their equations 
for determining velocity field have complicated form. Finally, 
in their formalism, 
it is necessary to solve a complicated vector Poisson equation for 
relativistic cases, which should be unnecessary for irrotational fluid. 
Although we may get correct results using their formalism, 
we had better obtain a simpler and more tractable formalism. 
The purpose in this paper is to present such one. 

In section II, we derive hydrostatic equations for 
irrotational fluid from relativistic hydrodynamic equations. 
We use 3+1 formalism and project the 
hydrodynamic equations onto $\Sigma_t$. 
Then, we impose an irrotational condition on $\Sigma_t$, which 
agrees with the relativistic irrotational condition\cite{HE}. 
As a result of projection onto $\Sigma_t$, we obtain hydrostatic equations 
on $\Sigma_t$, and hence, they 
have suitable forms to be solved in numerical relativity. 
Also, in our formalism, we need to solve 
only one Poisson type equation for a scalar field for determination 
of vector field. In section III, taking Newtonian limit, we 
show that well-known Newtonian hydrostatic equations 
are derived from the present formalism. 
In section IV, we give first post-Newtonian hydrostatic equations 
for irrotational fluid as well as gravitational potentials to be solved. 
Section V is devoted to summary. 
Throughout this paper, $c$ denotes speed of light, and we 
use units in which gravitational constant is unity. 
We use units $c=1$ in section II for convenience and recover $c$ in 
sections III and IV. Latin and Greek indices denote three dimensional (3D) 
spatial components ($1-3$) 
and four dimensional (4D) components ($0-3$), respectively. 
As spatial coordinates, we use the Cartesian coordinates $x^k=(x^1,x^2,x^3)$.  

\section{Relativistic fluid equations in 3+1 formalism} 

Since we use 3+1 formalism in general relativity, 
we write the line element as 
\beqn
ds^2&=&g_{\mu\nu}dx^{\mu}dx^{\nu} \nonumber \\
&=&(-\alpha^2+\beta_k\beta^k)dt^2
+2\beta_i dx^idt+\gamma_{ij}dx^idx^j ,
\eeqn
where $g_{\mu\nu}$, $\alpha$, $\beta_i=\gamma_{ij}\beta^j$ 
and $\gamma_{ij}$ are 4D metric, the 
lapse function, shift vector, and 3D spatial metric 
respectively. Using the unit normal to 3D spatial hypersurface $\Sigma_t$, 
\beq
n^{\mu}=\biggl({1 \over \alpha}, -{\beta^i \over \alpha}\biggr)~~~~
{\rm and}~~~~
n_{\mu}=\biggl(- \alpha, 0,0,0\biggr), 
\eeq
$\gamma_{ij}$ is written as
\beq
\gamma_{\mu\nu}=g_{\mu\nu}+n_{\mu}n_{\nu}.
\eeq
Hereafter, we use $\nabla_{\mu}$ and $D_i$ as the covariant 
derivatives with respect to $g_{\mu\nu}$ and $\gamma_{ij}$, 
respectively. 

We assume the energy momentum tensor of the perfect fluid as 
\beq
T^{\mu\nu}=\rho \biggl[1+\varepsilon+{P \over \rho}\biggr] 
u^{\mu}u^{\nu} + Pg^{\mu\nu},
\eeq
where $\rho$, $\varepsilon$, $P$, and $u^{\mu}$ denote the
rest mass density, specific internal energy, pressure, and
four velocity, respectively. We assume polytropic  
equation of state $P=(\Gamma-1)\rho\varepsilon$, where $\Gamma=1+1/n$ and
$n$ is the polytropic index. From adiabatic condition, we also 
get $P=K\rho^{\Gamma}$, where $K$ is a constant. For the following, 
we define $h$ as 
\beqn
h&&=1+  \varepsilon+{P \over \rho} \nonumber \\
&&=1+{K \Gamma \over \Gamma-1}\rho^{\Gamma-1}
=1+\int {dP \over \rho}. 
\eeqn

{}From the conservation equation for the energy momentum tensor, 
\beq
\nabla_{\mu} T^{\mu}_{~\nu}=0,
\eeq
we get the hydrodynamic equation as 
\beq
u^{\mu}\nabla_{\mu} \tilde u_{\nu}+\nabla_{\nu} h =0,\label{beq}
\eeq
where $\tilde u_{\nu}=hu_{\nu}$, and we use 
the conservation equation for rest mass density as 
\beq
\nabla_{\mu} (\rho u^{\mu})=0. \label{cceq}
\eeq

To rewrite the hydrodynamic equation, we decompose $u^{\mu}$ as 
\beq
u^{\mu}=u^0( \ell^{\mu}+V^{\mu}),\label{veq}
\eeq
and assume that (1) $\ell^{\mu}$ is a timelike vector 
of its component $(1, \ell^{i})$, 
and (2) $V^{\mu}$ is a spatial vector, $V^{\mu} n_{\mu}=0$, 
i.e., $V^{\mu}=(0, V^i)$. 
By using $\ell^{\mu}$ and $V^{\mu}$, we get the following relations;
\beqn
\gamma_i^{~\nu} \ell^{\mu}\nabla_{\mu}\tilde u_{\nu}&=&
\gamma_i^{~\nu}\biggl[
{\cal L}_{\ell} \tilde u_{\nu}-\tilde u_{\mu} \nabla_{\nu} \ell^{\mu} 
\biggr]\nonumber \\
&=&\gamma_i^{~\nu}\biggl[{\cal L}_{\ell} \tilde u_{\nu} -
\tilde u_{\mu} \nabla_{\nu}\biggl( {u^{\mu} \over u^0}-V^{\mu}\biggr)
\biggr] \nonumber \\
&=& \gamma_i^{~\nu} \biggl[ {\cal L}_{\ell} \tilde u_{\nu}+ 
h \nabla_{\nu}\Bigl( {1 \over u^0} \Bigr)
+\tilde u_{\mu} \nabla_{\nu} V^{\mu}\biggr]\nonumber \\
&=& \gamma_i^{~\nu} {\cal L}_{\ell}\tilde u_{\nu}+
h D_i\Bigl( {1 \over u^0} \Bigr)+{} ^{(3)}\tilde u_{k} D_i V^{k} \nonumber \\
&& \hskip 2cm 
+n_{\sigma} \tilde u^{\sigma} \gamma_i^{~\nu} V^{\mu}\nabla_{\nu} n_{\mu},
\eeqn
and 
\beq
\gamma_i^{~\nu} V^{\mu}\nabla_{\mu}\tilde u_{\nu}
=V^kD_k{} ^{(3)}\tilde u_i
- n_{\sigma}\tilde u^{\sigma}\gamma_i^{~\nu} V^{\mu}\nabla_{\mu} n_{\nu},
\eeq
where ${\cal L}_{\ell}$ denotes the Lie derivative 
with respect to $\ell^{\mu}$ and $^{(3)}\tilde u_i$ is a 
spatial vector defined as $\gamma_i^{~k} \tilde u_k$. 
Using these relations, projection of Eq. (\ref{beq}) onto 
the 3D hypersurface $\Sigma_t$ becomes 
\beqn
&&u^0\biggl[ \gamma_i^{~\nu} {\cal L}_{\ell}\tilde u_{\nu}+
V^kD_k{} ^{(3)}\tilde u_i+{} 
^{(3)}\tilde u_k D_i V^k+h D_i \Bigl({1 \over u^0}\Bigr)
 \biggr]\nonumber \\
&&\hskip 1cm +D_i h=0. 
\eeqn
We can rewrite this equation as 
\beqn
\gamma_i^{~\nu} {\cal L}_{\ell} \tilde u_{\nu}&+&
D_i \biggl({h \over u^0} +{}  ^{(3)}\tilde u_k V^k \biggr) \nonumber \\
&+& V^k (D_k{} ^{(3)}\tilde u_i-D_i{} ^{(3)}\tilde u_k)=0.\label{bbeq}
\eeqn

Besides the conservation equation of the energy momentum tensor, we 
have the conservation equation for rest mass density (\ref{cceq}). 
We note that for the case of barotropic equation of state 
such as $P=K\rho^{\Gamma}$, Eq. (\ref{cceq}) is also derived from the 
conservation equation of the energy momentum tensor. This implies that 
if we solve Eq. (\ref{cceq}), we do not have to take into account 
$n^{\mu}$ component of Eq. (\ref{beq}). 
Using Eq. (\ref{veq}), Eq. (\ref{cceq}) is written as 
\beq
\alpha[{\cal L}_{\ell} (\rho u^0)+ \rho u^0 \nabla_{\mu} \ell^{\mu}]+
D_i (\rho \alpha u^0 V^i)=0. \label{ceq}
\eeq

Now, we assume that $\ell^{\mu}$ is a Killing vector such that 
$\nabla_{\mu} \ell_{\nu}+\nabla_{\nu} \ell_{\mu}=0$, 
${\cal L}_{\ell} \tilde u_{\nu}=0$ and ${\cal L}_{\ell} (\rho u^0)=0$, 
and we write its component as $(1, -\Omega x^2, \Omega x^1,0)$, 
where $\Omega$ is identified with the orbital angular velocity 
with respect to distant inertial observer. 
We note that fluid exists inside the light cylinder 
$|x^k| \ll  c\Omega^{-1}$, 
and existence of the Killing vector is 
assumed within it. We also note that $\ell^{\mu}$ defined here 
is identical with the helicoidal Killing vector 
defined by BGM\cite{BGM}. If the Killing vector exists, 
we can derive hydrostatic equations 
for two interesting cases. One is the corotational case where 
we simply set $V^i=0$. Then, we get a well-known result as\cite{LPPT} 
\beq
{h \over u^0}={\rm constant},\label{coroteq}
\eeq
and continuity equation is trivially satisfied in this case. 

The other is the case where $^{(3)}\tilde u_i$ 
satisfies an ``irrotational condition'' defined as 
\beq
W_{ij} \equiv 
D_i {} ^{(3)}\tilde u_j - D_j{} ^{(3)}\tilde u_i=0, \label{irrecon}
\eeq
and hence 
\beq
^{(3)}\tilde u_i = D_i \phi, 
\eeq
where $\phi$ is a scalar field. 
Then, the hydrodynamic equation (\ref{bbeq}) is integrated to give 
\beq
{h \over u^0} +{}  ^{(3)}\tilde u_k V^k = {\rm constant}. \label{feq}
\eeq
Note that $V^k$ and $u^0$ are written as
\beqn
V^k&=&-\ell^k-\beta^k+{1 \over h u^0}\gamma^{kl} D_l \phi,\label{peq}\\
u^0&=&{1 \over \alpha}\biggl[1+ h^{-2}\gamma^{kl} D_k \phi D_l \phi 
\biggr]^{1/2},
\eeqn
so that we can rewrite left-hand side of Eq. (\ref{feq}) as 
\beq
{h \over u^0} +{}  ^{(3)}\tilde u_k V^k=
h\alpha^2 u^0-(\ell^k+\beta^k){} ^{(3)}\tilde u_k.\label{feq2}
\eeq

By substituting Eq. (\ref{peq}) into Eq. (\ref{ceq}), we get a 
Poisson type equation for determining $\phi$ as 
\beq
D_i (\rho \alpha h^{-1} D^i \phi)-
D_i\{\rho \alpha u^0(\ell^i+\beta^i)\}=0.\label{conteq}
\eeq
Hence, hydrodynamic equations which should be solved for 
determination of equilibrium states reduce to only two 
hydrostatic equations (\ref{feq}) and (\ref{conteq}). 
We do not have to solve any equations for vector potentials 
which were introduced in the formalism of BGM\cite{BGM}. 

We note that definition of irrotation 
in the 4D covariant form should be\cite{HE} 
\beqn
\omega_{\mu\nu}=&&P_{\sigma}^{~\mu}P_{\lambda}^{~\nu}(
\nabla_{\mu} u_{\nu}-\nabla_{\nu} u_{\mu}) \nonumber \\
=&&h^{-1}(\nabla_{\mu}\tilde u_{\nu}-\nabla_{\nu}\tilde  u_{\mu})=0,
\label{irreeq}
\eeqn
where $P_{\sigma}^{~\mu}=g_{\sigma}^{~\mu}+u_{\sigma} u^{\mu}$, and we 
use Eq. (\ref{beq}) to rewrite the first line into the second line. 
When $\omega_{\mu\nu}$ is vanishing initially for a fluid element, 
it remains zero along the trajectory of the fluid element 
for the perfect fluid\cite{HE}. 
Hence, $\omega_{\mu\nu}=0$ is just the irrotational condition. 
In our present irrotational condition (\ref{irrecon}), 
Eq. (\ref{irreeq}) is satisfied on the 3D hypersurface $\Sigma_t$ 
trivially. 
However, it is not trivial whether or not projection of 
Eq. (\ref{irreeq}) to $n^{\mu}\gamma^{\nu}_{~k}$ component 
is satisfied. (Projection to $n^{\mu}n^{\nu}$ component 
is trivially satisfied.) 
We here show that it is really guaranteed due to Eq. (\ref{irrecon}). 
By operating  $n^{\mu}\gamma^{\nu}_{~k}$ to 
$\nabla_{\mu}\tilde u_{\nu}-\nabla_{\nu}\tilde  u_{\mu}$, we get
\beqn
&&n^{\mu}\gamma^{\nu}_{~k}(\nabla_{\mu}\tilde u_{\nu}
-\nabla_{\nu}\tilde  u_{\mu}) \nonumber \\
=&&\gamma^{\nu}_{~k}{\cal L}_n \tilde u_{\nu} - 
\gamma^{\nu}_{~k}( \tilde u_{\mu}\nabla_{\nu} n^{\mu}
+n^{\mu}\nabla_{\nu} \tilde u_{\mu}) \nonumber \\
=&&\gamma^{\nu}_{~k}{\cal L}_{n}\tilde u_{\nu}- D_k ( n^{\mu} \tilde u_{\mu})
\nonumber \\
=&&\gamma^{\nu}_{~k}{\cal L}_{n}\tilde u_{\nu}+ D_k (h\alpha u^0)
\equiv W_k,
\eeqn
where we use $n^{\mu} \tilde u_{\mu}=-h\alpha u^0$. 
{}From a straightforward calculation, we can rewrite 
$\gamma^{\nu}_{~k} {\cal L}_n \tilde u_{\nu}$ as 
\beqn
\gamma^{\nu}_{~k} {\cal L}_n \tilde u_{\nu}={1 \over \alpha}\biggl[
\gamma^{\nu}_{~k}{\cal L}_{\ell} \tilde u_{\nu}
&&+h\alpha u^0 D_k \alpha
-(\beta^j+\ell^j) \pa_j{} ^{(3)}\tilde u_k \nonumber \\
&&- {} ^{(3)}\tilde u_j \pa_k (\beta^j +\ell^j)\biggr],
\eeqn
where $\pa_k$ denotes partial derivative on $\Sigma_t$. Hence, 
\beqn
W_k &&={1 \over \alpha} \biggl[\gamma^{\nu}_{~k} {\cal L}_{\ell} \tilde u_{\nu}
-(\beta^j+\ell^j) \pa_j{} ^{(3)}\tilde u_k 
\nonumber \\
&&~~~~~~~~-{} ^{(3)}\tilde u_j \pa_k (\beta^j +\ell^j)+\pa_k(h\alpha^2 u^0)
\biggr]. 
\eeqn
Using the hydrodynamic equation (\ref{bbeq}) and 
an identity (\ref{feq2}), we obtain
\beqn
W_k&=&{1 \over \alpha}(V^j+\beta^j+\ell^j)
(-\pa_j{} ^{(3)}\tilde u_k+\pa_k{} ^{(3)}\tilde u_j) \nonumber \\
&=&{1 \over \alpha}(V^j+\beta^j+\ell^j)W_{kj}.\label{qqqeq}
\eeqn
Eq. (\ref{qqqeq}) implies that 
$W_k=0$ if Eq. (\ref{irrecon}) is satisfied. 
Note that to derive Eq. (\ref{qqqeq}) we have not assumed the fact 
that $\ell^{\mu}$ is a Killing vector. 
Therefore, Eq. (\ref{irrecon}) is the necessary and sufficient condition 
for the irrotational condition in general case. 
Note that Eq. (\ref{feq}) itself does not mean irrotation in general. 
Even for the case when a Killing vector $\ell^{\mu}$ exists, 
it is nothing but a necessary condition for irrotation. 

\section{Newtonian limit}

In the Newtonian limit, metric variables can be expanded as 
\beqn
&&\alpha= 1 - {U \over c^2} + O(c^{-4}),\\
&&\beta^k =O(c^{-3}),\\
&&\gamma_{ij}=\delta_{ij}+ O(c^{-2}),
\eeqn
where $U$ denotes the Newtonian potential which satisfies 
\beq
\Delta U =-4\pi \rho,\label{pnpu}
\eeq
and $\Delta$ is the flat Laplacian. By using $v^i \equiv u^i/u^0$, 
components of $u^{\mu}$ which we need here is also expanded as 
\beqn
u^0&=&1+{1 \over c^2}\Bigl\{{1 \over 2}v^2+U \Bigr\}+O(c^{-4}),\\
u^i&=&u_i={v^i \over c}+O(c^{-3}),
\eeqn
where $v^2=\sum_i v^i v^i$. Note also $\ell^{\mu}=(1, \ell^i/c)$ and 
$V^{\mu}=(0, V^i/c)$. For the corotational case ($V^i=0$), we get 
Newtonian limit of left-hand side of Eq. (\ref{coroteq}) as 
\beq
\biggl[{h \over u^0}\biggr]_{\rm full~rela.} 
\longrightarrow 1+{1 \over c^2}\biggl[
-{v^2 \over 2}-U + \int {dP \over \rho}\biggr].\label{nceq}
\eeq
Since $V^k=0$, $v^k$ is equal to $\ell^k$, and $v^2=R^2 \Omega^2$ where 
$R^2=(x^1)^2+(x^2)^2$. Substituting this relation of $v^2$ 
into Eq. (\ref{nceq}), we get a well-known result
\beq
-{R^2 \Omega^2 \over 2}-U + \int {dP \over \rho} = {\rm constant}. 
\eeq

For the irrotational case, Newtonian limit of the 
left-hand side of Eq. (\ref{feq}) becomes
\beqn
&&\biggl[ {h \over u^0} +{}  ^{(3)}\tilde u_k V^k \biggr]_{\rm full~rela.} 
\longrightarrow \nonumber \\
&& 1+{1 \over c^2}\biggl[
-{v^2 \over 2}-U + \int {dP \over \rho}+\sum_k v^k (-\ell^k+v^k)\biggr]. 
\eeqn
In Newtonian order, $v^k = \pa_k \phi_{\rm N}$, where 
$\phi$ is expanded as $\phi_{\rm N}/c + O(c^{-3})$. So that we get 
\beq
{1  \over 2}\sum_k (\pa_k \phi_{\rm N})^2 - U + \int {dP \over \rho}-
\sum _k \ell^k \pa_k \phi_{\rm N} = {\rm constant}.\label{nieq}
\eeq
Eq. (\ref{nieq}) agrees with that of BGM\cite{BGM}. 

{}From continuity equations in Newtonian order, we obtain 
equations for $\phi_{\rm N}$ as 
\beq
\rho \Delta \phi_{\rm N} + \sum_k (\pa_k \phi_{\rm N} 
- \ell^k) \pa_k \rho =0. \label{conneq}
\eeq
Eq. (\ref{conneq}) is solved under boundary condition, 
\beq
\sum_k (\pa_k \phi_{\rm N}- \ell^k) \pa_k \rho =0, 
\eeq
at stellar surface.

\section{First post-Newtonian equations} 

In this section, we derive hydrostatic equations 
in first post-Newtonian order. The equations 
for the corotational case agree with those shown 
in previous papers\cite{chandra}\cite{shiba}, so that 
we here derive equations only for the irrotational 
case. In first post-Newtonian approximation, metric 
in the standard post-Newtonian gauge can be expanded as\cite{BDS} 
\beqn
&&\alpha= 1 - {U \over c^2} + {1 \over c^4}\biggl[ {U^2 \over 2} + X \biggr]
+O(c^{-6}),\\
&& \beta^k = {1 \over c^3}\hat \beta_k + O(c^{-5}), \\
&& \gamma_{ij}=\delta_{ij} \biggl[ 1+ {2 \over c^2} U \biggr]+ O(c^{-4}), 
\eeqn
where $X$ and $\hat \beta_k$ are obtained from 
\beqn
&&~~~ \Delta X = 4\pi\rho \biggl( 2U + 2 \sum_k (\pa_k \phi_{\rm N})^2
+\varepsilon +{3P \over \rho} \biggr), \label{pnp1} \\
&&~~~\hat \beta_k =  -{7 \over 2}P_k
+{1 \over 2} \biggl( \pa_k \chi + \sum_j x^j \pa_k P_j \biggr),\\
&&{\rm and} \nonumber \\
&&~~~ \Delta P_k = -4 \pi \rho \pa_k \phi_{\rm N}, \label{pnp2} \\
&&~~~ \Delta \chi=4 \pi \rho \sum_k (\pa_k \phi_{\rm N}) x^k.\label{pnp3} 
\eeqn
Note that to derive these Poisson equations, we use a relation 
in Newtonian order, $v^k=\pa_k \phi_{\rm N}$. 

Using a post-Newtonian relation, 
\beqn
\alpha u^0 = 1&+&{1 \over 2c^2}\sum_k (\pa_k \phi_{\rm N})^2
+{1 \over c^4}\biggl[
-{1 \over 8}\Bigl(\sum_k (\pa_k \phi_{\rm N})^2\Bigr)^2 \nonumber \\
&+&\sum_k \pa_k \phi_{\rm N} \pa_k \phi_{\rm PN}
-(\eta+U) \sum_k (\pa_k \phi_{\rm N})^2 \biggr] \nonumber \\
&+& O(c^{-6}), 
\eeqn
where we expand $\phi$ as $\phi_{\rm N}/c + \phi_{\rm PN}/c^3+O(c^{-5})$ 
and $\eta=\varepsilon +P/\rho$, 
first post-Newtonian expansion of Eq. (\ref{feq}) becomes 
\beqn
&\biggl[& {h \over u^0} +{}  ^{(3)}\tilde u_k V^k \biggr]_{\rm full~rela.}
\longrightarrow \nonumber \\
&1&+{1 \over c^2}\biggl[\eta-U
+{1 \over 2}\sum_k (\pa_k \phi_{\rm N})^2 - \sum_k \ell^k \pa_k \phi_{\rm N}
\biggr] 
\nonumber \\
&~&+{1 \over c^4}\biggl[-\eta U +{1 \over 2}U^2 + X 
-{1 \over 2}(\eta + 3U)\sum_k (\pa_k \phi_{\rm N})^2 \nonumber \\
&&\hskip 1cm -{1 \over 8}\Bigl( \sum_k (\pa_k \phi_{\rm N})^2 \Bigr)^2 
 +\sum_k \pa_k \phi_{\rm N} \pa_k \phi_{\rm PN} \nonumber \\
&& \hskip 10mm 
-\sum_k \ell^k \pa_k \phi_{\rm PN}- \sum_k \hat \beta_k \pa_k \phi_{\rm N}
\biggr]
\nonumber \\
&=&{\rm constant}. \label{pneq1}
\eeqn
First post-Newtonian expansion of continuity equation is also derived as  
\beq
\sum_i \pa_i (\rho A_i)=0,\label{pneq2}
\eeq
where 
\beqn
A_i=&& -\ell^i + \pa_i \phi_{\rm N}
+{1 \over c^2}\biggl\{ -\ell^i\biggl({1 \over 2}\sum_k (\pa_k \phi_{\rm N})^2
+3U\biggr) \nonumber \\
&&-\eta \pa_i \phi_{\rm N} -\hat \beta_i + \pa_i \phi_{\rm PN}
\biggr\}. 
\eeqn
If  $\phi_{\rm N}$ is obtained from Eq. (\ref{conneq}), 
Eq. (\ref{pneq2}) is regarded as equation for $\phi_{\rm PN}$ and 
solved under boundary condition 
\beq
\sum_i A_i \pa_i \rho = 0
\eeq
at stellar surface. 

Eqs. (\ref{pneq1}) and (\ref{pneq2}) with Poisson equtions 
(\ref{pnp1}), (\ref{pnp2}), (\ref{pnp3}) and (\ref{pnpu}) 
are basic equations for computation of 
irrotational equilibrium states in first post-Newtonian order. 

\section{Summary}

In this paper, we have derived 
relativistic hydrostatic equations for obtaining 
irrotational (quasi) equilibrium configurations 
of binary neutron stars using 3+1 formalism. 
In order to derive the hydrostatic equations, 
we first projected hydrodynamic equations onto $\Sigma_t$ 
and then impose the irrotational condition to 
obtain the hydrostatic equations. 
As a result, the hydrostatic equations obtained are simple and 
suitable for numerical relativity, 
compared with a previous formalism\cite{BGM}. Also, 
as a natural consequence, in our formalism there is 
no vector Poisson equation to be solved and 
only a scalar Poisson equation is needed to be solved for determination of 
velocity field not only for Newtonian, but also for relativistic cases. 

We also give hydrostatic equations as well as Poisson equations for 
gravitational potentials needed for computation of irrotational 
equilirium states in first post-Newtonian approximation. We think that 
as a first step toward fully relativistic study, 
we had better construct post-Newtonian configurations for 
firm investigation of relativistic effect on binary 
neutron stars. In reality, we have been able to obtain 
much information on relativistic effect in binary 
neutron stars from first post-Newtonian studies\cite{shiba}\cite{tani}
\cite{shibata}. Up to now, however, our 
attention was only paid to corotational binary neutron stars. 
The present formalism makes it possible to extend previous studies to 
the irrotational case. As a first work, we plan to obtain 
incompressible, irrotational equilibrium states of binary stars 
as we carried out for the corotational case previously\cite{tani}. 

The author thanks M. Sasaki, T. Tanaka  and H. Asada for 
helpful discussion. This work is in part supported by a 
Japanese Grant-in-Aid of 
Ministry of Education, Culture, Science and Sports 
(Nos. 08NP0801 and 09740336).


\begin{thebibliography}{99}

\bibitem{LIGO} A. Abramovici et al. Science {\bf 256}, 325 (1992).
\bibitem{VIRGO}
C. Bradaschia, et al. Nucl. Instrum. and Methods {\bf A289}, 518 (1990).
\bibitem{GEO}
J. Hough, in {\it Proceedings of the Sixth Marcel Grossmann Meeting},
edited by H. Sato and T. Nakamura
(World Scientific, Singapore, 1992), p.192.
\bibitem{TAMA}
K. Kuroda et al., in {\it Proceedings of the international conference on
gravitational waves: Sources and Detectors}, edited by
I. Ciufolini and F. Fidecard(World Scientific, 1997), p.100. 

\bibitem{KIP} For example, K. S. Thorne, in {\it 
Proceeding of Snowmass 94 Summer Study on Particle and Nuclear 
Astrophysics and Cosmology}, eds. E. W. Kolb and R. Peccei 
(World Scientific, Singapore, 1995) and references cited therein. 

\bibitem{lindblom}
L. Lindblom, Astrophys. J. {\bf 398}, 569 (1992).

\bibitem{shiba} M. Shibata, Phys. Rev. D {\bf 55}, 6019 (1997);
Prog. Theor. Phys. {\bf 96}, 317 (1996); 
M. Shibata, K. Taniguchi, and T. Nakamura,
Prog. Theor. Phys. supplement {\bf 128}, 295 (1997).

\bibitem{BCSST} 
T. W. Baumgarte, G. B. Cook, M. A. Scheel, S. L. Shapiro, 
and S. A. Teukolsky,
Phys. Rev. Lett. {\bf 79}, 1182 (1997); gr-qc/9709026.

\bibitem{uryu} However, for Newtonian case, following researchers 
have obtained irrotational Newtonian equilibrium states; \\
K. Uryu and Y. Eriguchi, astro-ph/9712203; \\
S. Bonazzola. E. Gourgoulhon, J.-A. Marck, talk at GRG 15, India (1997). 

\bibitem{Cutler} 
C. S. Kochanek, Astrophys. J. {\bf 398}, 234(1992);\\
L. Bildsten and C. Cutler, Astrophys. J. {\bf 400}, 175(1992).

\bibitem{BGM}
S. Bonazzola. E. Gourgoulhon, J.-A. Marck, Phys. Rev. D {\bf 56}, 7740 
(1997).

\bibitem{HE}
S. W. Hawking and G. F. R. Ellis, {\it The large scale structure of 
space-time} (Cambridge University Press, 1973), 84. 


\bibitem{LPPT} A. P. Lightman, W. H. Press, R. H. Price and 
S. A Teukolsky, {\it Problem Book in Relativity and Gravitation} 
(Princeton Unversity Press, 1975), problem 16.17. 

\bibitem{chandra}
S. Chandrasekhar, Astrophys. J. {\bf 148}, 621 (1967). 


\bibitem{BDS} S. Chandrasekhar, Astrophys. J. {\bf 142}, 1488 (1965): \\
L. Blanchet, T. Damour, and G. Sch\"afer, 
Mon. Not. R. astr. Soc. {\bf 242}, 289 (1990). 

\bibitem{tani} K. Taniguchi and M. Shibata, Phys. Rev. D {\bf 56}, 798
(1997). 

\bibitem{shibata} M. Shibata and K. Taniguchi  Phys. Rev. D {\bf 56}, 811
(1997). 
\end{thebibliography}
\end{document}